\documentclass{iopart}
\usepackage{graphicx}  
\usepackage{latexsym}  

\jl{6}        
\eqnobysec    

\def\beq{\begin{equation}}
\def\eeq{\end{equation}}
\def\rmd{{\rm d}}

\begin{document}

\title[Spinning particles in the vacuum C metric]
{Spinning particles in the vacuum C metric}

\author{
Donato Bini$^* {}^\S{}^\P$,
Christian Cherubini$^\dagger {}^\S$, 
Andrea Geralico$^\ddag {}^\S$,
and Bahram Mashhoon$^\diamond$}
\address{
  ${}^*$\
Istituto per le Applicazioni del Calcolo ``M. Picone'', CNR I-00161 Rome, Italy
}
\address{
  ${}^\S$\
  International Center for Relativistic Astrophysics,
  University of Rome, I-00185 Rome, Italy
}
\address{
${}^\P$
  INFN - Sezione di Firenze, Polo Scientifico, Via Sansone 1, 
  I-50019, Sesto Fiorentino (FI), Italy 
}
\address{
${}^\dagger$\
Facolt\`a di Ingegneria, Universit\`a ``Campus Biomedico'', Via E. Longoni 47, I-00155 Rome, Italy}

\address{
  ${}^\ddag$\
  Dipartimento di Fisica, Universit\`a di Lecce, and INFN - Sezione di Lecce,
  Via Arnesano, CP 193, I-73100 Lecce, Italy}

\address{${}^\diamond$
Department of Physics and Astronomy, University of Missouri-Columbia, Columbia, Missouri 65211, USA
}

\begin{abstract}
   The motion of a spinning test particle given by the Mathisson-Papapetrou equations is studied on an exterior vacuum C metric background spacetime describing the accelerated motion of a spherically symmetric gravitational source. We consider circular orbits of the particle around the direction of acceleration of the source. The symmetries of this configuration lead to the reduction of the differential equations of motion to algebraic relations. The spin supplementary conditions as well as the coupling between the spin of the particle and the acceleration of the source are discussed.
\end{abstract}

\pacno{04.20.Cv}

\section{Introduction}

In a recent paper \cite{bcm}, the absence of spin-acceleration coupling has been examined via wave perturbations of the vacuum C metric 
\cite{kin69,kinwal,Farh,ES,bcmprd}, which represents the static exterior gravitational field of a spherical mass $M$ that is accelerated uniformly with acceleration $A$ such that $MA < 1/(3\sqrt{3})$. This gravitational field may be considered to be a nonlinear superposition of the Rindler and Schwarzschild spacetimes. The aim of the present work is to investigate the motion of a classical spinning test particle in the C metric using the Mathisson-Papapetrou equations \cite{math37,papa51}. For the sake of simplicity, we confine our analysis to motion along circular orbits about the direction of acceleration. We find that for a given circular orbit the results are unchanged when the particle spin flips and at the same time the sense of motion reverses, a circumstance that is consistent with time-reversal invariance. Our results thus indicate that a spin-acceleration coupling analogous to the spin-rotation coupling \cite{mash74,mash88} does not exist. This particle result thus reinforces and complements the previous wave result \cite{bcm}.
  
This paper is organized as follows: In section 2, we present a brief 
discussion of the geometric properties of circular orbits around the 
direction of acceleration of the C metric. In section 3, we discuss the 
motion of spinning test particles along the circular orbits for the 
three standard spin supplementary conditions and determine the allowed 
orbits and spin directions. Section 4 contains a discussion of our 
results.  

\section{Vacuum C metric and circular orbits}

In its restricted interpretation, the vacuum C metric describes the static gravitational field associated with a
 uniformly accelerated mass with $MA<1/(3\sqrt{3})$ \cite{kin69,kinwal}; it is of Petrov type D and belongs to the Weyl class of solutions of the Einstein equations \cite{ES}.
In the $\{u,r,\theta,\phi \}$ coordinate system the C metric takes the form
\begin{equation}
\label{cmetu}
\fl\qquad
\rmd s^2= -H \rmd u^2 - 2 \rmd u \rmd r + 2A r^2 \sin \theta \rmd u \rmd \theta +\frac{r^2\sin^2\theta }{G} \rmd \theta^2 +r^2 G \rmd \phi^2  ,
\end{equation} 
where $G$ and $H$ are given by
\begin{eqnarray}
\fl\qquad
G(\theta)&=& 1-\cos^2\theta -2MA \cos^3\theta , \nonumber \\
{}\fl\qquad
H(r,\theta)&=&1-\frac{2M}{r}-A^2r^2 (1-\cos^2\theta-2MA\cos^3\theta) \nonumber \\
{}\fl\qquad
&&-2Ar\cos\theta(1+3MA\cos\theta)+6MA\cos\theta .
\end{eqnarray}

The constants $M> 0$ and $A> 0$ denote the mass and acceleration of the source, respectively. 
Unless specified otherwise, we choose units such that the gravitational constant and the speed of light in vacuum are unity.
Moreover, we assume that the C metric has signature +2; to preserve this signature, we must have $G>0$. As mentioned above,  it turns out that the physical region of interest in this case corresponds to $MA<1/(3\sqrt{3})$ \cite{Farh,pavda,podol} .

The metric (\ref{cmetu}) can be seen to be a nonlinear superposition of two metrics, one  associated with a Schwarzschild black hole (case $A=0$) and the other corresponding to a uniformly accelerating particle 
(case $M=0$) \cite{kin69,kinwal}. 
The C metric has event horizons (which are also Killing horizons) given by hypersurfaces of the form $r=r(\theta)$ that are solutions of $H=0$. These can be determined exactly.
To this end, let us introduce the new variable \cite{bcmprd}
\beq
W= \frac{Ar}{1-Ar\cos \theta} 
\eeq
so that $H=0$ becomes 
\beq
\label{eqW}
W^3-W+2MA=0.
\eeq 
Following \cite{bcmprd},  let us also introduce an acceleration lengthscale based on $A>0$:
$L_A=\frac{1}{3\sqrt{3}A}$.
There are three cases  depending on whether $M$ is less than, equal to or greater than $L_A$. As we are interested only in the case $M<L_A$,
 the solutions of (\ref{eqW}) can be written as
\beq
W_1=2{\hat U}, \qquad W_2=-{\hat U}+\sqrt{3}{\hat V}, \qquad W_3=-{\hat U}-\sqrt{3}{\hat V},
\eeq
where
\beq
{\hat U}+i{\hat V}=\frac{1}{\sqrt{3}} \left(-\frac{M}{L_A} +i \sqrt{1-\frac{M^2}{L_A^2}} \right)^{1/3}.
\eeq
It is worth noticing that, when $M\to 0$, the solution  $W_1$ gives the Rindler horizon $r=[A(1+\cos\theta)]^{-1}$, while, when $A\to 0$,
the solution  $W_2$ gives the Schwarzschild horizon $r=2M$; $W_3$ gives instead a negative value for $r$ and should be rejected.
Thus $r=[A(\cos\theta+1/W_1)]^{-1}$ and $r=[A(\cos\theta+1/W_2)]^{-1}$ correspond to the modified Rindler and Schwarzschild horizons, respectively.

The C metric has two commuting hypersurface-orthogonal Killing vectors
one timelike  ($\partial_u$) and the other spacelike ($\partial_\phi$).

\begin{figure}[h]
\typeout{*** EPS figure 1}
\begin{center}
\includegraphics[scale=0.4]{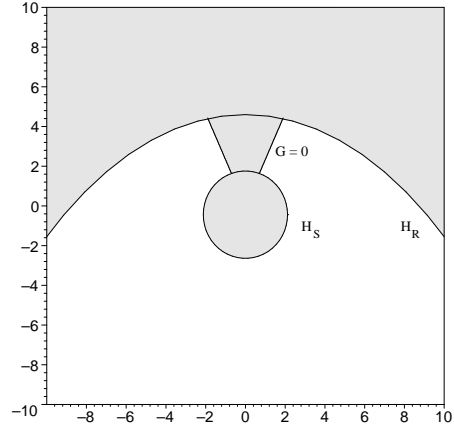}  
\end{center}
\caption{The accessible spacetime region (not shaded) in the $(X,Y)$ plane with $X={\hat r}\sin \theta$ (abscissa) and $Y={\hat r}\cos \theta$ (ordinate), ${\hat r}=r/M$,  is shown for the value of the parameter $MA=0.1$. 
The upper curve represents the Rindler horizon while the lower curve represents the Schwarzschild horizon. The forbidden conical region corresponds to negative values of the metric function $G$, i.e. to signature changes which are not considered here.}  
\label{fig:1}
\end{figure}

To illustrate this situation, in Figure \ref{fig:1} we have shown 
the accessible spacetime region in the ($r, \theta$) plane for the value of the metric parameter $MA=0.1$. 
In this plot, it is possible to identify both the Rindler horizon ($H_R$) and the Schwarzschild horizon ($H_S$) as well as the forbidden conical region corresponding to negative values of the metric function $G$, i.e. to signature changes.

In the spacetime represented by the metric (\ref{cmetu}), let us consider as \lq\lq fiducial observers"  the family of static observers with 4-velocity  aligned with the Killing direction $\partial_u$, $e_{\hat u}=H^{-1/2}\partial_u $;
it is convenient to introduce an orthonormal frame adapted to these observers
\begin{eqnarray}
\label{frame}
& e_{\hat u}=H^{-1/2}\partial_u\ , \quad 
e_{\hat r}=H^{-1/2}[H\partial_r-\partial_u]\ , \nonumber \\ 
& e_{\hat \theta}=G^{1/2}[Ar\partial_r+\frac1{r\sin\theta }\partial_{\theta}]\ , \quad 
e_{\hat \phi}=\frac{G^{-1/2}}{r}\partial_\phi\ .
\end{eqnarray}

We are interested in the study of circular motion of test particles along the $\phi$ direction with constant speed and at a fixed value of the polar angle $\theta$ as illustrated in Figure 2; so the family of test particles is characterized by the following (timelike) 4-velocity $U$
\begin{equation}
\label{circolare}
U=\Gamma_{\zeta}[\partial_u + \zeta\partial_{\phi}]=\gamma [e_{\hat u} +\nu  e_{\hat\phi}]\ ,
\end{equation}
where $\zeta$ and $\nu$ are the angular velocity and linear velocity  parametrizations, respectively, of the whole family such that
\begin{equation}
\nu=r\left(\frac{G}{H}\right)^{1/2} \zeta .
\end{equation}
Here $\Gamma_{\zeta}$ is defined by the timelike condition $U\cdot U=-1$ as
\begin{equation}
-\Gamma_{\zeta}^{-2}=g_{uu}+\zeta^2g_{\phi\phi}=-H+\zeta^2r^2G=-\frac{H}{\gamma^2}\ 
\end{equation}
and $\gamma=-U\cdot e_{\hat u}=(1-\nu^2)^{-1/2}=\Gamma_\zeta H^{-1/2}$ is the Lorentz factor.
Note that $\Gamma_{\zeta}$, $\zeta$, $\nu$ and $\gamma$  are all functions of $r$ and $\theta$ only; therefore, they remain constant along the orbit (see Figure 2). 

\begin{figure}[h]
\typeout{*** EPS figure 2}
\begin{center}
\includegraphics[scale=0.4]{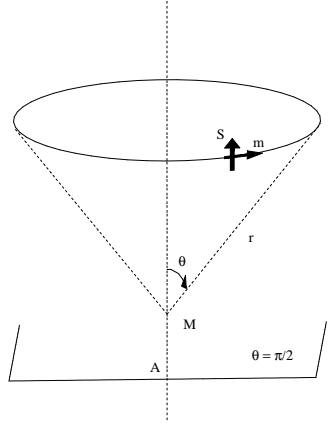}   
\end{center}
\caption{Schematic diagram representing circular orbits under consideration in this paper. The orbital plane is parallel to the $\theta=\pi/2$ plane. The spherical source $M$ is accelerated with uniform acceleration $A$ in the $\theta=\pi$ direction.}  
\label{fig:2} 
\end{figure}

We also define the unit direction orthogonal to $U$ in the $(e_{\hat u} , e_{\hat \phi})$ 2-space
\beq
E_{\hat \phi}= \gamma [\nu e_{\hat u} +  e_{\hat\phi}],
\eeq
which is a vector field defined only along the world line of $U$; in fact, its definition requires the relative velocity field $\nu$ of $U$ with respect to fiducial observers $e_{\hat u}$ which only exist along $U$. Hereafter, the 1-form field associated with $E_{\hat \phi}$ will be denoted as $\Omega^{\hat \phi}$ and the standard notation $X^\flat$ for the completely covariant form of a tensor $X$ will be adopted. In this sense,  $E_{\hat \phi}^\flat=\Omega^{\hat \phi}$.

Two quantities are relevant for the discussion below: the acceleration of the line $U$, 
$a(U)= D U/{\rmd \tau_U }=\nabla_U U $, where $\tau_U$ is the proper time along $U$,
and the directional derivative along $U$ of the local $\phi$ direction as seen by the observer $U$, $D E_{\hat \phi}/{\rmd \tau_U }=\nabla_U E_{\hat \phi}$.
A direct evaluation shows that both these quantities belong to the ($e_{\hat r}, e_{\hat \theta}$) plane and are given by

\begin{eqnarray}\label{acccomp_gen}
\frac{D U}{\rmd \tau_U }  &=&\gamma^2 \left[ k{}_{\hat r} (\nu^2-\nu_{(r)}^2) e_{\hat r}+ k{}_{\hat \theta} (\nu^2-\nu_{(\theta )}^2) e_{\hat \theta}\right],\\
\label{acccomp_gen2}
\frac{D E_{\hat \phi}}{\rmd \tau_U }&=& \gamma^2 \nu \left(\frac{k{}_{\hat r}}{\gamma_{(r)}^2} e_{\hat r} + \frac{k{}_{\hat \theta}}{\gamma_{(\theta)}^2} e_{\hat \theta} \right),
\end{eqnarray}
where
\begin{eqnarray}
\label{nurthdef}
\fl\quad
\nu_{(r)}^2&=& \frac{M-Ar^2[GAr+\cos\theta(1+3AM\cos\theta)]}{rH}, \quad \gamma_{(r)}^2=1/(1-\nu_{(r)}^2)\ ,\nonumber \\
\fl\quad
\nu_{(\theta )}^2&=& \frac{GAr}{GAr+\cos\theta(1+3AM\cos\theta)} , \quad \gamma_{(\theta)}^2=1/(1-\nu_{(\theta)}^2)\ ,
\end{eqnarray}
and the Lie relative curvature \cite{idcf1,idcf2} of the orbit $k_{\rm (Lie)}\equiv k= -\nabla\ln{(\sqrt{g_{\phi\phi}})}$ has been introduced 
with frame components
\begin{equation}
\label{liecurv}
\fl\quad
k{}_{\hat r} =-\frac{\sqrt{H}}{r}\ , 
\qquad  k{}_{\hat \theta } = -\frac{GAr+\cos\theta(1+3AM\cos\theta)}{r\sqrt{G}}\ ,
\end{equation}
and magnitude $ \kappa=(k{}_{\hat r}^2 + k{}_{\hat \theta}^2)^{1/2} $.

Analogously, it is easy to derive the directional derivatives along $U$ of $e_{\hat r}$ and $e_{\hat \theta}$, which instead belong to the $(e_{\hat u}, e_{\hat \phi})$ plane:
\begin{eqnarray}\label{eretheta}
\frac{D e_{\hat r}}{\rmd \tau_U }  &=&\gamma^2 k{}_{\hat r}\left[  (\nu^2-\nu_{(r)}^2) U- \frac{\nu}{\gamma_{(r)}^2}E_{\hat \phi}\right],\\
\frac{D e_{\hat \theta}}{\rmd \tau_U }&=& \gamma^2 k{}_{\hat \theta}\left[  (\nu^2-\nu_{(\theta)}^2) U- \frac{\nu}{\gamma_{(\theta)}^2}E_{\hat \phi}\right] .
\end{eqnarray}

The following useful relations  hold
\beq
k{}_{\hat r}^2 \nu_{(r)} ^2 + k{}_{\hat \theta }^2 \nu_{(\theta )}^2  =\frac{M}{r^3}, \quad A \sqrt{G}=-k{}_{\hat \theta} \nu_{(\theta)}^2\ .
\eeq
A detailed analysis of circular orbits, \lq\lq special" for geometrical or kinematical reasons, is considered in \cite{bcgj}. 

\begin{figure} [h]
\typeout{*** EPS figure 3}
\begin{center}
\includegraphics[scale=0.4]{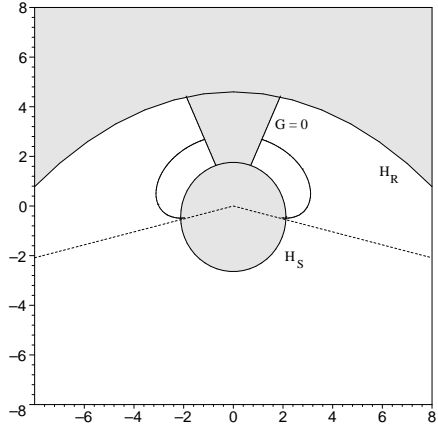}  
\end{center}
\caption{Circular geodesics exist for fixed values of $\theta$ only for certain corresponding values of $r$. 
The solid line in the allowed region represents the location of the geodesics, which are respectively timelike, null and spacelike above, on and below the equatorial plane $\theta=\pi/2$.  The pair of straight lines dropping down from the center correspond to the limiting value $\theta=\theta_c$. The value of the parameter $MA=0.1$ is the same as in Figure \ref{fig:1}.} 
\label{fig:3}
\end{figure}

The only non-vanishing components of the 4-acceleration $a(U)=\nabla_U U$  
can be conveniently rewritten as
\begin{eqnarray}\label{comp_acc}
\fl\quad
a(U)^{\hat r}      =\gamma^2 k{}_{\hat r} (\nu^2-\nu_{(r)}^2)\ , \qquad 
a(U)^{\hat \theta }=\gamma^2 k{}_{\hat \theta} (\nu^2-\nu_{(\theta )}^2)\ .
\end{eqnarray}
Conditions for circular geodesics, namely $(a(U)^{\hat r},a(U)^{\hat \theta })=(0,0)$ (or equivalently, $\nu=\pm \nu_{(r)}$ and $\nu_{(r)}^2= \nu_{(\theta)}^2$) can be easily obtained from (\ref{comp_acc}); it follows that for a fixed  angle $\theta$, only the value $r=r_g$ is allowed with
\begin{eqnarray}
\fl\quad
\label{rgeo}
r_g&=&\frac{1}{2A}\frac{3AMG+(AM)^{1/2}[\cos\theta(4+AM\cos^3\theta)+9AM(3-2G)]^{1/2}}{G+\cos\theta[\cos\theta+3AM(2-AM\cos^3\theta)]}\ ,
\end{eqnarray}
where
\begin{eqnarray}\fl\quad
\label{nugeo}
\nu \equiv \nu_{g\pm}=  \pm \sqrt{\frac{Ar_gG}{Ar_gG+\cos\theta(1+3MA\cos\theta)}}\ , \qquad \gamma_{g}=1/\sqrt{1-\nu_{g\pm}^2}\ .
\end{eqnarray}
Note that $r_g$ and $ \nu_{g\pm}$ depend on $\theta$ only. 
In the same diagram of Figure \ref{fig:1}, the curve of the $r$ and $\theta$ values corresponding to circular geodesics is shown in Figure \ref{fig:3}.

The circular geodesics of the C metric, 
which represents the gravitational potential of a Schwarzschild black hole 
accelerating uniformly with acceleration $A < 1/ (3\sqrt{3} M)$ 
along the $\theta = \pi$ direction, have been obtained with the assumption that 
$A > 0$; moreover, the geodesics are timelike for $\theta_0 < \theta < \pi/2$ 
and spacelike for $\pi/2 < \theta < \theta_c$. 
There is a unique null geodesic for $\theta = \pi/2$ given by $r_g = 3 M$ 
irrespective of of the acceleration $A$ so long as $MA < 1/(3\sqrt{3})$. 
Here $\theta_0$ and $\theta_c$  depend upon $MA$, $\theta_0$ is given by $G(\theta_0)=0$ 
and $\theta_c$ is such that $r_g (\theta_c)$ lies on the modified 
Schwarzschild horizon that satisfies $H ( r_g(\theta_c), \theta_c ) = 0$, 
see Figure 3. 
In the nonrelativistic limit $MA << 1$, 
the circular geodesics reduce to the orbits that can be easily 
obtained from the Newtonian theory of gravitation: $( M/r_g^2) \cos \theta = A$ 
follows from (\ref{rgeo}) and $\nu_g^2 = (M/r_g) \sin^2 \theta$ 
from (\ref{rgeo}) - (\ref{nugeo}). Furthermore, in the Newtonian limit $\zeta_g^2 = M/r_g^3$. 

It is interesting to consider the significance of these orbits 
in the gravitoelectromagnetic analogy \cite{bcmprd}. 
When one explores the correspondence between classical and 
quantum descriptions of electron motion in the Stark effect, 
the circular orbits that correspond to those under 
consideration here describe the classical motion of the 
electron in a quantum-mechanical phenomenon, namely, 
the {\it quadratic} Stark effect in hydrogen 
\cite{hooker}. 

The purpose of this paper is to explore the accelerated 
motion of classical spinning test particles on circular orbits $r = r( \theta)$ 
in the C metric. 
We therefore turn to the investigation of the allowed orbits for particles with spin.

\section{Spinning test particles along circular orbits}

The motion of a (classical) spinning test particle is described by the Mathisson-Papapetrou equations
\begin{eqnarray}
\label{papcoreqs1}
\frac{DP^{\mu}}{\rmd \tau_U}&=&-\frac12R^{\mu}{}_{\nu\alpha\beta}U^{\nu}S^{\alpha\beta}\equiv F_{\rm spin}^{\mu} ,\\
\label{papcoreqs2}
\frac{DS^{\mu\nu}}{\rmd \tau_U}&=&P^{\mu}U^{\nu}-P^{\nu}U^{\mu}\ ,
\end{eqnarray}
where $P^{\mu}$ is the total 4-momentum of the particle and $S^{\mu\nu}$ is the antisymmetric spin tensor. Here $U$ is the timelike unit tangent vector of the \lq\lq center line'' that is used to perform the multipole reduction of the energy-momentum tensor associated with the particle in order to derive equations (\ref{papcoreqs1}) and (\ref{papcoreqs2}).

We assume here the center line to be a circular orbit, i.e. $U$ is given by (\ref{circolare}).
From (\ref{papcoreqs2}), contracting both sides of the second equation with $U_\nu$, one has
\begin{equation}
\label{Ps}
P^{\mu}=-(U\cdot P)U^\mu -U_\nu \frac{DS^{\mu\nu}}{\rmd \tau_U}\equiv
mU^\mu +P_s^\mu ,
\end{equation}
where $m$ is the component of $-P$ along $U$ and it is usually referred to as the mass of the particle under the Pirani supplementary conditions (see below).
Equation (\ref{papcoreqs2}) implies that
\begin{equation}\fl\quad
\label{spinconds}
S_{\hat u\hat \phi}=0\ , \quad 
S_{\hat r\hat \theta }=0\ , \quad 
k{}_{\hat r} \left( \nu_{(r)}^2S_{{\hat \theta }{\hat u}}+\nu S_{{\hat \theta }{\hat \phi}}\right)
-k{}_{\hat \theta } \left(\nu_{(\theta)}^2S_{{\hat r}{\hat u}} +\nu S_{{\hat r}{\hat \phi}}\right)=0\ . 
\end{equation}
The spin tensor then takes the form
\begin{equation}
\label{Sform}
S=\omega^{\hat r}\wedge [S_{\hat r\hat u}\omega^{\hat u}+S_{\hat r\hat \phi}\omega^{\hat \phi}]+\omega^{\hat \theta }\wedge [S_{\hat \theta \hat u}\omega^{\hat u}+S_{\hat \theta \hat \phi}\omega^{\hat \phi}]\ ,
\end{equation}
where $\{\omega^{\hat u},\omega^{\hat r},\omega^{\hat \theta},\omega^{\hat \phi} \}$ is the dual frame of (\ref{frame}).
It is clear from (\ref{Ps}) that $P_s$ is orthogonal to $U$; moreover, it turns out to be also aligned with  $E_{\hat \phi}$  
\begin{equation}
\label{ps}
P_s=m_s E_{\hat \phi} \ ,
\end{equation}
where 
\begin{eqnarray}
\label{msdef}
\fl\quad
m_s\equiv||P_s||=\gamma\left[k{}_{\hat r} \left( \nu S_{{\hat r}{\hat u}}+\nu_{(r)}^2 S_{{\hat r}{\hat \phi}}\right)+
k{}_{\hat \theta} \left(\nu S_{{\hat \theta}{\hat u}} +\nu_{(\theta)}^2 S_{{\hat \theta}{\hat \phi}}\right)
\right]
\end{eqnarray}
is the component of $P$ along $E_{\hat \phi}$.
It follows from (\ref{Ps}) and (\ref{ps}) that the total 4-momentum $P$  can be written in the form $P=\mu \, U_p$, where 
$\mu$ represents the magnitude of $P$ and the unit timelike direction $U_p$ is given by
\begin{equation}
\label{Ptot}
\fl\quad
U_p=\gamma_p\, [e_{\hat u}+\nu_p e_{\hat \phi}]\ , \quad \nu_p=\frac{\nu+m_s/m}{1+\nu m_s/m}\ , \quad  \mu=\frac{\gamma}{\gamma_p}(m+\nu m_s)\ .\ 
\end{equation}
Here $\gamma_p=(1-\nu_p^2)^{-1/2}$, as expected. 

Let us now consider the equation of motion (\ref{papcoreqs1}). 
The spin force is given by:
\begin{eqnarray}
\label{Fspin}
F^{\rm (spin)}=\gamma\frac{M}{r^3}\left[ (2S_{{\hat u}{\hat r}}+\nu  S_{{\hat r}{\hat \phi}})e_{\hat r}
-(S_{{\hat u}{\hat \theta }}+2\nu S_{{\hat \theta}{\hat \phi}})e_{\hat \theta}\right]\ ,
\end{eqnarray}
while the term on the left-hand side of equation (\ref{papcoreqs1}) can be written, from (\ref{Ps}) and (\ref{ps}),  as 
\beq
\label{motrad}
\frac{DP}{\rmd \tau_U}=m a(U)+m_s \frac{DE_{\hat \phi}}{\rmd \tau_U}\ ,
\eeq
where $a(U)$ and $DE_{\hat \phi}/{\rmd \tau_U}$ are given by (\ref{acccomp_gen}) and (\ref{acccomp_gen2}), respectively, and the quantities  $\mu$, $m$ and $m_s$ remain constant along the world line of $U$. 
Hence (\ref{papcoreqs1}) can be written as
\begin{eqnarray}\fl\quad
\label{eqmotonew}
0&=&\gamma\left[m(\nu^2-\nu_{(r)}^2)+m_s\frac{\nu}{\gamma_{(r)}^2}\right]k{}_{\hat r}-\frac{M}{r^3}(2S_{{\hat u}{\hat r}}+\nu  S_{{\hat r}{\hat \phi}}),\nonumber\\
\fl\quad
0&=& \gamma\left[m(\nu^2-\nu_{(\theta)}^2)+m_s\frac{\nu}{\gamma_{(\theta)}^2}\right]k{}_{\hat \theta} +\frac{M}{r^3}(S_{{\hat u}{\hat \theta }}+2\nu S_{{\hat \theta}{\hat \phi}})\ .
\end{eqnarray}

Finally, it proves useful to introduce  the quadratic invariant 
\beq
\label{sinv}
s^2=\frac12 S_{\mu\nu}S^{\mu\nu}=-S_{\hat r\hat u }^2-S_{\hat \theta \hat u }^2+S_{\hat r \hat \phi}^2+S_{\hat \theta \hat \phi}^2\ , 
\eeq
which is not in general constant along the trajectory of the spinning particle. It is constant, however, in the situation under consideration in this paper.

Summarizing, the equations to be considered are: 
a) the constraint equations (\ref{spinconds}) that result from the evolution of the spin tensor whose nontrivial components are given in (\ref{Sform});
b) the two equations of  motion (\ref{eqmotonew}), c) the definition of $m_s$ given in (\ref{msdef}) and d) the quadratic invariant $s$ introduced in (\ref{sinv}).
To solve these equations and, hence, to discuss the features of the motion we need three further supplementary conditions. 
We shall follow the standard approaches existing in the literature:
\begin{itemize}
\item[1.]
Corinaldesi-Papapetrou \cite{cori51} conditions (CP): $S^{u\nu}=0$,
\item[2.]
Pirani \cite{pir56} conditions (P): $S^{\mu\nu}U_\nu=0$, 
\item[3.]
Tulczyjew \cite{tulc59} conditions (T): $S^{\mu\nu}P_\nu=0$.
\end{itemize}

It is helpful to introduce the spin vector
\beq
S_{(SC)}{}_\beta=\frac12 \eta^\alpha{}_{\beta\gamma\delta}(U_{(SC)})_\alpha S^{\gamma\delta},
\eeq
where $U_{(SC)}=(e_{\hat u},U,U_p)$ for the CP, P, T conditions, respectively, and 
$\eta_{\alpha\beta\gamma\delta}=\sqrt{-g} \epsilon_{\alpha\beta\gamma\delta}$ is the unit volume 4-form constructed with the Levi-Civita alternating symbol $\epsilon_{\alpha\beta\gamma\delta}$, such that $\epsilon_{\hat u\hat r\hat\theta\hat\phi}=1$. It follows that in general $s^2=S_{\mu}S^{\mu}$.

A consistent physical interpretation of the Mathisson-Papapetrou
equations requires that ${\hat s}:= s/ (m r ) \ll 1$. For $s = 0$, the orbit of
the test particle coincides with the geodesic orbit $r_g = r_g (\theta_g)$; therefore, we expect that the circular orbit $r = r (\theta )$ of the
spinning particle should be close to such a geodesic. Hence, we assume that 
\beq
r=r_g+\delta_r\ , \qquad \theta=\theta_g+\delta_{\theta}\  ,
\eeq 
where $\delta_r$ and $\delta_{\theta}$ are of order $\hat s$ and will only be considered to linear order in what follows.
Any given function of the particle orbit, $F(r,\theta)$, can then be written  to first order in  $\delta_{r}$ and $\delta_{\theta}$ as
\beq
F(r,\theta)=F(r_g , \theta_g )+ \left(\frac{\partial F}{\partial r} \right)_{r=r_g,\, \theta=\theta_g} \delta_r +
 \left(\frac{\partial F}{\partial \theta} \right)_{r=r_g,\, \theta=\theta_g} \delta_\theta .
\eeq
In the following, we shall obtain for each SC explicit expressions in terms of $r$ and $\theta$ for both the linear velocity $\nu$ and the spin parameter $\hat s$ associated with the orbit of the spinning particle.
Close to a geodesic, we shall show that  all the SC cases can be summarized as
\begin{eqnarray}
\label{solSCexpnu}
\nu\simeq \nu_{g\pm}+ \sigma\Delta\nu{}^{(SC)},\qquad
\hat s \simeq  \sigma \Delta{\hat s}{}^{(SC)}\ ,
\end{eqnarray}
where the notation $\sigma={\rm sign }(\nu_{g \pm})$ is conveniently introduced, so that $\sigma=+1$ for a counterclockwise orbit and $\sigma=-1$ for a clockwise orbit (cf. Figure 2). 
The corresponding angular velocity $\zeta$ and its reciprocal are
\begin{eqnarray}
\label{solSCzetamen1}
\zeta_{\pm}\simeq\zeta_{g\pm}+\sigma\Delta\zeta{}^{(SC)}\ , \qquad
\frac1{\zeta}_{\pm}\simeq \frac1{\zeta_{g\pm}}-\frac{\sigma}{\zeta_{g\pm}^2}\Delta\zeta{}^{(SC)} , \ 
\end{eqnarray}
respectively, with $\zeta_{g\pm}=(\nu_{g\pm}\sqrt{H_g})/(r_g\sqrt{G_g})$.  
As a result, the orbital frequency turns out to depend on the spin parameter.
Explicit expressions for all of the quantities introduced above can be found in Appendix A.

\subsection{The Corinaldesi-Papapetrou (CP) supplementary conditions}

These are {\it conceptually} the weakest conditions that have been imposed on the spin tensor; in general, they do not even ensure that the quadratic invariant $S_{\mu\nu} S^{\mu\nu}$ is conserved along the path. In the simple configuration under consideration here, the CP supplementary conditions require $S_{\hat u \hat r}=S_{\hat u \hat \theta}=0$.
Equations (\ref{spinconds}) as well as (\ref{sinv}) are identically satisfied with
\beq
(S_{{\hat r}{\hat \phi}}, S_{{\hat \theta}{\hat \phi}})=\frac{s}{\kappa}(k{}_{\hat r},k{}_{\hat \theta}),
\eeq
where $s$ can be of any sign.
The  spin vector,
\beq
S_{(CP)}{}_\beta=\frac12 \eta^\alpha{}_{\beta\gamma\delta}(e_{\hat u})_\alpha S^{\gamma\delta},
\eeq
spatial with respect to $e_{\hat u}$ in this case is given by
\beq
S_{(CP)}=-S_{{\hat r}{\hat \phi}} e_{\hat \theta}+S_{{\hat \theta}{\hat \phi}} e_{\hat r}=\frac{s}{\kappa}[-k{}_{\hat r} e_{\hat \theta}+
k{}_{\hat \theta} e_{\hat r}].
\eeq
It follows, from (\ref{msdef}) that 
\beq
m_s=\frac{M}{r^3}\frac{s\gamma}{\kappa}\ .
\eeq
Finally, the spin force (see (\ref{Fspin})) is given by 
\beq
F^{(\rm spin)}=\frac{s\gamma\nu}{\kappa} \frac{M}{r^3} [k{}_{\hat r} e_{\hat r}-2k{}_{\hat \theta} e_{\hat \theta}]\ ,
\eeq
so that equations (\ref{eqmotonew}) reduce to
\begin{eqnarray}
\label{ssolCP12}
0&=& (\nu^2-\nu_{(r)}^2)\left(s+\frac{m}{\gamma \nu}\kappa\frac{r^3}{M}\right)\ , \nonumber\\
0&=&s+ \left(\frac{m}{\gamma \nu}\kappa\frac{r^3}{M} \right )\frac{\nu^2-\nu_{(\theta )}^2}{3-2\nu^2-\nu_{(\theta )}^2}\ .
\end{eqnarray}
Now, by eliminating $s$, we get the solution for $\nu=\nu^s_{\pm}$. 
If $\nu^2\not =\nu_{(r)}^2$, then plugging the first of equations (\ref{ssolCP12}) into the second, one gets $\nu^s_\pm=\pm 1$ and so $s=0$, which must be rejected. The only possibility is then $\nu =\nu_{(r)}$ (see (\ref{nurthdef})), with  $s$ given by
\beq
s= - \frac{m}{\gamma_{(r)} \nu_{(r)}}\kappa\frac{r^3}{M} \frac{
\nu_{(r)}^2-\nu_{(\theta )}^2}{3-2\nu_{(r)}^2-\nu_{(\theta )}^2}\ .
\eeq

Close to a geodesic, from (\ref{ssolCP12}) we obtain
\begin{eqnarray}\fl\quad
\label{ssolCP12exp}
\Delta\nu{}^{(CP)}&=&\Delta_{\nu_{(r)}}^r\delta_r + \Delta_{\nu_{(r)}}^{\theta}\delta_{\theta}\ , \nonumber\\
\fl\quad
\Delta{\hat s}{}^{(CP)}&=&-\frac23\gamma_{g}\kappa_{g}\frac{r_g^2}{M}\left[\delta_r(\Delta_{\nu_{(r)}}^r-\Delta_{\nu_{(\theta)}}^r)+\delta_{\theta}(\Delta_{\nu_{(r)}}^{\theta}-\Delta_{\nu_{(\theta)}}^{\theta})\right]\ ;
\end{eqnarray}
the linear velocity $\nu$ and the corresponding angular velocity $\zeta$ and its reciprocal are thus given by (\ref{solSCexpnu}) - (\ref{solSCzetamen1}), respectively.

\subsection{The Pirani (P) supplementary conditions}

It was first demonstrated by Pirani \cite{pir56} that the Mathisson-Papapetrou equations with the supplementary conditions $S^{\mu\nu}U_{\nu} = 0$ imply that the spin vector is Fermi-Walker transported along the trajectory. Subsequently, Taub \cite{taub} demonstrated that these equations correspond to the motion of a classical point particle with \lq\lq intrinsic" spin. The Mathisson-Papapetrou equations were investigated in \cite{masspin1,masspin2} and it was found that of the various supplementary conditions the only one that produces a consistent result in the massless limit is $S^{\mu\nu} U_{\nu} = 0$ \cite{masspin2}. In fact, the trajectory in this case is a null geodesic with the spin vector either parallel or antiparallel to the direction of motion; furthermore, the helicity of the particle is conserved in an orientable spacetime \cite{masspin2}.

In the case under consideration here, the P supplementary conditions ($S^{\mu\nu}U_\nu=0$) require 

\beq
S_{\hat u\hat r}=S_{\hat r \hat \phi}\nu\ , \qquad
S_{\hat u\hat \theta }=S_{\hat \theta \hat \phi}\nu\ .
\eeq
Using these relations one finds that  equations (\ref{spinconds}) as well (\ref{sinv}) are identically satisfied with
\beq
(S_{{\hat r}{\hat \phi}}, S_{{\hat \theta}{\hat \phi}})=\frac{\gamma s}{\Sigma}\left(\frac{k{}_{\hat r}}{\gamma_{(r)}^2},\frac{k{}_{\hat \theta}}{\gamma_{(\theta)}^2}\right),
\eeq
where 
\beq
\Sigma =\left(\frac{k{}_{\hat r}^2}{\gamma_{(r)}^4} +\frac{k{}_{\hat \theta}^2}{\gamma_{(\theta)}^4}\right)^{1/2}
\eeq
and, as before, $s$ can be of any sign.
The spin vector
\beq
S_{(P)}{}_\beta=\frac12 \eta^\alpha{}_{\beta\gamma\delta}U_\alpha S^{\gamma\delta}
\eeq
spatial with respect to $U$ in this case is given by
\beq
S_{(P)}=\frac{1}{\gamma}[-S_{{\hat r}{\hat \phi}} e_{\hat \theta}+S_{{\hat \theta}{\hat \phi}} e_{\hat r}]=
\frac{s}{\Sigma}\left[-\frac{k{}_{\hat r}}{\gamma_{(r)}^2} e_{\hat \theta}+
\frac{k{}_{\hat \theta}}{\gamma_{(\theta) }^2} e_{\hat r}
\right].
\eeq

Then, from (\ref{msdef}) we have that 
\beq
m_s=-s\gamma^2 \frac{\gamma_{(\theta)}^2 k{}_{\hat r}^2 (\nu^2-\nu_{(r)}^2) + \gamma_{(r)}^2 k{}_{\hat \theta}^2 (\nu^2-\nu_{(\theta)}^2)}{\gamma_{(r)}^2 \gamma_{(\theta)}^2 \Sigma} .
\eeq
Finally, the spin force (see (\ref{Fspin})) is given by 
\beq
F^{(\rm spin)}=
\frac{3 s\gamma^2 \nu}{\Sigma} 
\frac{M}{r^3} 
\left( 
\frac{
k{}_{\hat r}}{ \gamma_{(r)}^2}e_{\hat r} -
\frac{k{}_{\hat \theta}}{ \gamma_{(\theta)}^2}e_{\hat \theta}
 \right),
\eeq
so that equations (\ref{eqmotonew}) reduce to
\begin{eqnarray}
\label{ssolP12}
s&=& -\frac{m \gamma_{(r)}^2 \Sigma}{\nu}\frac{\nu^2-\nu_{(r)}^2}{(\kappa^2-\frac{4M}{r^3})-\gamma^2\Sigma^2},
\nonumber\\
s&=&- \frac{m\gamma_{(\theta)}^2\Sigma }{\nu}\frac{\nu^2-\nu_{(\theta)}^2}{(\kappa^2+\frac{2M}{r^3})-\gamma^2\Sigma^2}, 
\end{eqnarray}
by solving both equations with respect to $s$.
By eliminating $s$ in (\ref{ssolP12}), we get the solution $\nu=\nu^s_{\pm}$:
\begin{eqnarray}
\label{nusolP}
\nu^s_{\pm}&=& 
\pm 
\left[ 
1-\frac{\frac{6M}{r^3}-\Sigma^2 (\gamma_{(\theta)}^2-\gamma_{(r)}^2)}{(\kappa^2+\frac{2M}{r^3})\gamma_{(r)}^2-(\kappa^2-\frac{4M}{r^3})\gamma_{(\theta)}^2}
\right]^{1/2}.
\end{eqnarray}

Close to a geodesic,  we obtain from (\ref{ssolP12}) -  (\ref{nusolP}) that
\begin{eqnarray}\fl\quad
\label{ssolP12exp}
\Delta\nu{}^{(P)}&=&\frac13\bigg\{\frac{\kappa_{g}^2r_g^3}{2M}\nu_{g\pm}^2[(\Delta_{\nu_{(r)}}^r-\Delta_{\nu_{(\theta)}}^r)\delta_r+(\Delta_{\nu_{(r)}}^{\theta}-\Delta_{\nu_{(\theta)}}^{\theta})\delta_{\theta}]+ (\Delta_{\nu_{(r)}}^r+2\Delta_{\nu_{(\theta)}}^r)\delta_r\nonumber\\
\fl\quad
&&+(\Delta_{\nu_{(r)}}^{\theta}+2\Delta_{\nu_{(\theta)}}^{\theta})\delta_{\theta}\bigg\}\ , \nonumber\\
\fl\quad
\Delta{\hat s}{}^{(P)}&=&\frac{\kappa_{g}}{3}\frac{r_g^2}{M}\left[\delta_r(\Delta_{\nu_{(r)}}^r-\Delta_{\nu_{(\theta)}}^r)+\delta_{\theta}(\Delta_{\nu_{(r)}}^{\theta}-\Delta_{\nu_{(\theta)}}^{\theta})\right]\ ; \, \,
\end{eqnarray}
the linear velocity $\nu$ and the corresponding angular velocity $\zeta$ and its reciprocal are given by (\ref{solSCexpnu}) - (\ref{solSCzetamen1}), respectively.

\subsection{The Tulczyjew (T) supplementary conditions}

The general theory of the motion of an extended particle in a gravitational field has been developed by Dixon (see \cite{dixon} and references therein) on the basis of the dynamical equations $\nabla_\nu T^{\mu\nu} = 0$. In Dixon's approach, when attention is limited to a pole-dipole particle and higher multipole moments of the extended body are neglected, one recovers the Mathisson-Papapetrou equations with the supplementary conditions  $S^{\mu\nu} P_{\nu} = 0$. These require, in the case under consideration here, that  

\beq
S_{\hat u\hat r }=S_{\hat r \hat \phi}\nu_p\ , \qquad
S_{\hat u\hat \theta }=S_{\hat \theta \hat \phi}\nu_p\ .
\eeq
Using these relations one finds that  equations (\ref{spinconds}) as well as (\ref{sinv}) are identically satisfied with
\beq
(S_{{\hat r}{\hat \phi}}, S_{{\hat \theta}{\hat \phi}})=\frac{\gamma_p s}{\Lambda}
\left(k{}_{\hat r}(\nu-\nu_p\nu_{(r)}^2),k{}_{\hat \theta}(\nu-\nu_p\nu_{(\theta)}^2)
\right),
\eeq
where
\beq
\Lambda=\left[ k{}_{\hat r}^2(\nu-\nu_p\nu_{(r)}^2)^2 + k{}_{\hat \theta }^2(\nu-\nu_p\nu_{(\theta )}^2)^2\right]^{1/2}\ .
\eeq
The spin vector
\beq
S_{(T)}{}_\beta=\frac12 \eta^\alpha{}_{\beta\gamma\delta}U_p{}_\alpha S^{\gamma\delta}
\eeq
spatial with respect to $U_p$ in this case is given by
\beq
\fl\quad
S_{(T)}=\frac{1}{\gamma_p}[-S_{{\hat r}{\hat \phi}} e_{\hat \theta}+S_{{\hat \theta}{\hat \phi}} e_{\hat r}]=
\frac{s}{\Lambda}[-k{}_{\hat r} (\nu-\nu_p\nu_{(r)}^2) e_{\hat \theta}+
k{}_{\hat \theta}(\nu-\nu_p\nu_{(\theta)}^2) e_{\hat r}
].
\eeq
Moreover, from (\ref{msdef}) we have that 
\beq\fl\quad
m_s=-\frac{s\gamma\gamma_p}{\Lambda}[k{}_{\hat r}^2(\nu\nu_p-\nu_{(r)}^2)(\nu-\nu_p\nu_{(r)}^2)+ k{}_{\hat \theta }^2(\nu\nu_p-\nu_{(\theta )}^2)(\nu-\nu_p\nu_{(\theta )}^2)]\ ,
\eeq
and  the spin force (see (\ref{Fspin})) is given by 
\beq\fl\quad
F^{(\rm spin)}=\frac{s\gamma\gamma_p}{\Lambda}\frac{M}{r^3}[(\nu+2\nu_p)(\nu-\nu_p\nu_{(r)}^2)k{}_{\hat r}e_{\hat r}-(2\nu+\nu_p)(\nu-\nu_p\nu_{(\theta )}^2)k{}_{\hat \theta }e_{\hat \theta }]\ .
\eeq
Now, by solving both equations (\ref{eqmotonew}) with respect to $s$, we get
\begin{eqnarray}
\label{ssolT}
s&=&-m\frac{\gamma}{\gamma_p}\frac{\nu^2-\nu_{(r)}^2}{\frac{\gamma^2}{\gamma_{(r)}^2}\nu\tilde m_s+\frac{M}{\Lambda r^3}(2\nu_p+\nu)(\nu_p\nu_{(r)}^2-\nu)},\nonumber\\
s&=&-\frac{m}{\gamma\gamma_p\gamma_{(\theta)}^2}\frac{\gamma^2-\gamma_{(\theta)}^2}{\frac{\gamma^2}{\gamma_{(\theta)}^2}\nu\tilde m_s-\frac{M}{\Lambda r^3}(\nu_p+2\nu)(\nu_p\nu_{(\theta)}^2-\nu)}\ ,
\end{eqnarray}
where the quantity $\tilde m_s$ stands for $\tilde m_s=m_s/(s\gamma\gamma_p)$; by eliminating $s$, we have that $\nu_p$ must satisfy the following equation:
\beq
\label{eqnuP}
{\mathcal A}\nu_p^2+{\mathcal B}\nu_p+{\mathcal C}=0\ ,
\eeq
with
\begin{eqnarray}
\phantom{+}{\mathcal A}&=& \frac{3M}{r^3} \nu_{(r)}^2 (\nu^2-\nu_{(\theta)}^2)\ ,
\nonumber\\
-\frac{{\mathcal B}}{\nu}&=&(\nu_{(r)}^2-\nu_{(\theta)}^2)[(\nu^2+\nu_{(r)}^4)k{}_{\hat r}^2+(\nu^2+\nu_{(\theta)}^4)k{}_{\hat \theta}^2]\nonumber\\
&&-\frac{M}{r^3}[(\nu_{(r)}^2+2\nu_{(\theta)}^2)(1+\nu^2)-3(\nu^2+\nu_{(r)}^2\nu_{(\theta)}^2)]\ ,
\nonumber\\
\phantom{+}{\mathcal C}&=&-3\frac{M}{r^3}\nu^2(\nu^2-\nu_{(r)}^2)\ .
\end{eqnarray}
Let $\nu_p^{(\pm)}=(-{\mathcal B}\pm\sqrt{\Delta})/(2{\mathcal A})$ be the solutions of (\ref{eqnuP}). The reality condition for these solutions requires that $\nu$ take values outside the regions implicitly defined by the equation $\Delta\equiv {\mathcal B}^2-4{\mathcal A}{\mathcal C}=0$. By substituting $\nu_p=\nu_p^{(\pm)}$ into either equations (\ref{ssolT}), we obtain a relation between $\nu$ and $s$,
which must be considered together with the following further equation directly resulting from the definition (\ref{Ptot}) of $\nu_p$:
\beq
\label{ssolTfromnup}
s=-m\frac{\nu-\nu_p}{\gamma\gamma_p{\tilde m_s}(1-\nu\nu_p)}\bigg\vert_{\nu_p=\nu_p^{(\pm)}}\ .
\eeq
As a result, solutions for both quantities $\nu^s_{\pm}$ and $s$ can be derived explicitly from (\ref{ssolT}) and (\ref{ssolTfromnup}).

Close to a geodesic we obtain the same results as in the case of P supplementary conditions with
\beq
\nu_p^{(\pm)}=\nu+O({\hat s}^2)\ ;
\eeq
so the linear velocity $\nu$ and the corresponding angular velocity $\zeta$ and its reciprocal are given by (\ref{solSCexpnu}) - (\ref{solSCzetamen1}), respectively, with $\Delta\nu^{(T)}\equiv\Delta\nu^{(P)}$.

\section{Discussion}

The test-particle approximation in the case of the Mathisson-Papapetrou equations for a spinning particle implies that the acceleration of the world line due to the spin force defined by (\ref{papcoreqs1}) should be relatively small. This means in the case under consideration that qualitatively
\begin{eqnarray}
\label{cond41}
\fl\quad
{\rm Spin \, force} / {\rm Newtonian\,  force} \sim \frac{[( GM/r^3 ) s/c ]}{ [GMm/r^2
]}  =\frac{ s}{ ( m c r ) }\ll 1,                   
\end{eqnarray}
since in the multipole expansion around $U$, the dipole term should be smaller than the monopole term;
we neglect quadrupole and higher-order terms in (\ref{papcoreqs1}) and (\ref{papcoreqs2}).
The condition (\ref{cond41}) can be expressed as $r_M\ll r$, where $r_M$ is the M\o ller radius of the test mass \cite{mol}.

It follows from this condition that spinning test particles can be maintained on circular orbits in the vacuum C metric only for allowed values of $r$ and $\theta$ that are close to those of geodesic circular orbits that have been discussed in section 2 for spinless test particles. Thus we have considered a one-parameter family of solutions in each of the CP, P and T cases. That is, the spinning particle follows a circular orbit such that $r - r_g$ and $\theta - \theta_g$ are linear in $\hat s$ and higher-order terms in $\hat s$ have been neglected. We find that the orbital frequency is in general spin-dependent, but there is no clock effect in contrast to the Schwarzschild case \cite{bdfg1}, since for a given orbit, the orbital frequency is unchanged under the transformation $\nu \to - \nu$ and $\hat s \to -\hat s$. 
In fact, it follows from (\ref{solSCzetamen1}) that the difference in the arrival times after one complete revolution with respect to a static observer vanishes 
\beq\label{deltat}
\Delta t_{(+,+;-,-)}= 2\pi \left(\frac{1}{\zeta_{(SC,+,+)}}+\frac{1}{\zeta_{(SC,-,-)}}\right)=0\ ,
\eeq
where $\zeta_{(SC,\pm,\pm)}$ denotes the angular velocity of $U$ derived under a particular choice of supplementary conditions and corresponding to a $\sigma=\pm$ orbit with spin-up/down alignment. This is a consequence of the fact that the linear velocity $\nu$ and the spin parameter $\hat s$ must have both the same sign as $\nu_{g\pm}$, see (\ref{solSCexpnu}). 
In the Schwarzschild case, instead, when the motion is confined to the equatorial plane the Mathisson-Papapetrou equations give rise to a unique relation between $\nu$ and $\hat s$, which becomes linear 
in the limit of small values of the spin parameter.
All combinations of the relative signs between rotation and the two spin-up and spin-down orientations along the $z$-axis are  allowed, and so a non-zero clock effect can appear.

One may consider this circumstance intuitively as follows: In the case of the Schwarzschild circular equatorial orbits of a spinning test particle with its spin normal to the equatorial plane, one may imagine a transformation of coordinates to the frame comoving with the test particle. In this frame, the Schwarzschild source follows circular GEODESIC orbits about the spinning test particle; therefore, one would qualitatively expect the existence of a gravitomagnetic clock effect proportional to the specific angular momentum of the test particle. This effect would then be  expected to occur in the original coordinate system as well. It is interesting to observe that a  similar argument for the case of the C metric would fail, since the orbits of the source in the frame comoving with the test particle would not be geodesics by assumption.

\appendix

\section{Expansion of kinematical quantities around a geodesic}

The expansion of the metric functions $G$ and $H$, the linear velocities $\nu_{(r)}$ and $\nu_{(\theta)}$ and the components of the Lie curvature close to a selected geodesic can be expressed as  
\begin{eqnarray}
\fl\quad
G&\simeq&G_g+2A\sin\theta_g\frac{G_gr_g}{\nu_{g\pm}^2\gamma_{g\pm}^2}\delta_{\theta}
\equiv G_g+\Delta_G^{\theta}\delta_{\theta}\ , \nonumber\\
\fl\quad
H&\simeq&H_g+2\frac{\nu_{g\pm}^2H_g}{r_g}\delta_r+2A\sin\theta_g(1-Ar_g\cos\theta_g)[r_g-3M(1-Ar_g\cos\theta_g)]\delta_{\theta}\nonumber\\
\fl\quad
&&\equiv H_g+\Delta_H^r\delta_r+\Delta_H^{\theta}\delta_{\theta}\ , \nonumber\\
\fl\quad
k{}_{\hat r}&\simeq&-\frac{\sqrt{H_g}}{r_g}+\frac{\sqrt{H_g}}{\gamma_{g\pm}^2r_g^2}\delta_r-\frac{\Delta_H^\theta}{2r_g\sqrt{H_g}}\delta_{\theta}
\equiv k{}_{\hat r}^g+\Delta_{k{}_{\hat r}}^r\delta_r+ \Delta_{k{}_{\hat r}}^{\theta}\delta_{\theta}\ , \nonumber\\
\fl\quad
k{}_{\hat \theta}&\simeq&-\frac{A\sqrt{G_g}}{\nu_{g\pm}^2}+\frac{A\sqrt{G_g}}{\nu_{g\pm}^2\gamma_{g\pm}^2r_g}\delta_r+ \frac{1}{2r_g\sqrt{G_g}}\bigg[2\sin\theta_g(1+6AM\cos\theta_g)\nonumber\\
\fl\quad
&&+Ar_g\Delta_G^{\theta}\left(\frac1{\nu_{g\pm}^2}-2\right)\bigg]\delta_{\theta}
\equiv k{}_{\hat \theta}^g+\Delta_{k{}_{\hat \theta}}^r\delta_r+ \Delta_{k{}_{\hat \theta}}^{\theta}\delta_{\theta}\ , \nonumber\\
\fl\quad
\kappa&\simeq&(k{}_{\hat r}^g{}^2+k{}_{\hat \theta}^g{}^2)^{1/2}+\frac{k{}_{\hat r}^g\Delta_{k{}_{\hat r}}^r+k{}_{\hat \theta}^g\Delta_{k{}_{\hat \theta}}^{r}}{(k{}_{\hat r}^g{}^2+k{}_{\hat \theta}^g{}^2)^{1/2}}\delta_r+ \frac{k{}_{\hat r}^g\Delta_{k{}_{\hat r}}^{\theta}+k{}_{\hat \theta}^g\Delta_{k{}_{\hat \theta}}^{\theta}}{(k{}_{\hat r}^g{}^2+k{}_{\hat \theta}^g{}^2)^{1/2}}\delta_{\theta}\nonumber\\
\fl\quad
&&\equiv \kappa_g+\Delta_{\kappa}^r \delta_r+ \Delta_{\kappa}^{\theta}\delta_\theta\ , \nonumber\\
\fl\quad
\nu_{(r)}&\simeq&\nu_{g\pm}-\left[\frac{A^2G_gr_g}{H_g\nu_{g\pm}}\left(\frac1{\nu_{g\pm}^2}+\frac12\right)+\frac{\nu_{g\pm}}{2r_g}(1+2\nu_{g\pm}^2)\right]\delta_r+ \nu_{g\pm}\bigg\{\frac{A}{\Delta_H^r}\bigg[-Ar_g\Delta_G^{\theta}\nonumber\\
\fl\quad
&&+\sin\theta_g(1+6AM\cos\theta_g)\bigg]-\frac{\Delta_H^{\theta}}{2H_g}\bigg\}\delta_{\theta}
\equiv\nu_{g\pm}+\Delta_{\nu_{(r)}}^r\delta_r + \Delta_{\nu_{(r)}}^{\theta}\delta_{\theta}\ , \nonumber\\
\fl\quad
\nu_{(\theta)}&\simeq&\nu_{g\pm}+\frac{\nu_{g\pm}}{2\gamma_{g\pm}^2r_g}\delta_r+ \frac{\nu_{g\pm}}{2\gamma_{g\pm}^2}\left[\frac{\Delta_G^{\theta}}{G_g}+\frac{2\sin^2\theta_g}{\Delta_G^{\theta}}(1+6AM\cos\theta_g)\right]\delta_{\theta}\nonumber\\
\fl\quad
&&\equiv\nu_{g\pm}+\Delta_{\nu_{(\theta)}}^r\delta_r+ \Delta_{\nu_{(\theta)}}^{\theta}\delta_{\theta}\ , 
\end{eqnarray}
where $G_g=G(\theta_g)$ and $H_g=H(r_g, \theta_g)$. As a result, we obtain for each choice of SC the following approximate expressions for the linear velocity $\nu$ and the spin parameter $\hat s$:
\begin{eqnarray}
\nu&\simeq&\nu_{g\pm}+[\Delta_{\nu}^r{}^{(SC)}\delta_r + \Delta_{\nu}^{\theta}{}^{(SC)}\delta_{\theta}]
\equiv\nu_{g\pm}+\sigma\Delta\nu{}^{(SC)}\nonumber\\
\hat s&\simeq&[\Delta_{\hat s}^r{}^{(SC)}\delta_r + \Delta_{\hat s}^{\theta}{}^{(SC)}\delta_{\theta}]
\equiv\sigma\Delta{\hat s}{}^{(SC)}\ .
\end{eqnarray}
The corresponding angular velocity $\zeta$ and its reciprocal are then given by
\begin{eqnarray}\fl\quad
\zeta_{\pm}&\simeq&\zeta_{g\pm}+\sigma\frac{|\zeta_{g\pm}|}{|\nu_{g\pm}|}\bigg\{\Delta\nu^{(SC)}-\frac{|\nu_{g\pm}|}{2}\bigg[\frac{\delta_r}{r_g}\left(2-r_g\frac{\Delta_H^r}{H_g}\right)\nonumber\\
\fl\quad
&&+\delta_{\theta}\left(\frac{\Delta_G^{\theta}}{G_g}-\frac{\Delta_H^{\theta}}{H_g}\right)\bigg]\bigg\}
\equiv\zeta_{g\pm}+\sigma\Delta\zeta{}^{(SC)}\ 
\end{eqnarray}
and
\begin{eqnarray}\fl\quad
\frac1{\zeta_{\pm}}&\simeq&\frac1{\zeta_{g\pm}}-\frac{\sigma}{|\zeta_{g\pm}||\nu_{g\pm}|}\bigg\{\Delta\nu^{(SC)}-\frac{|\nu_{g\pm}|}{2}\bigg[\frac{\delta_r}{r_g}\left(2-r_g\frac{\Delta_H^r}{H_g}\right)\nonumber\\
\fl\quad
&&+\delta_{\theta}\left(\frac{\Delta_G^{\theta}}{G_g}-\frac{\Delta_H^{\theta}}{H_g}\right)\bigg]\bigg\} 
\equiv\frac1{\zeta_{g\pm}}-\frac{\sigma}{\zeta_{g\pm}^2}\Delta\zeta{}^{(SC)}\ .
\end{eqnarray}


\section*{References}
\begin{thebibliography}{00}

\bibitem{bcm} 
Bini D Cherubini C Mashhoon B 2004
{\it Class.\ Quantum Grav.\/} {\bf 21} 3893

\bibitem{kin69}
Kinnersley W 1969
{\it Phys.\, Rev.\/} {\bf 186} 1335 

\bibitem{kinwal}
Kinnersley W  and Walker M 1970 
{\it   Phys.\, Rev. D\/ } {\bf 2} 1359

\bibitem{Farh}
Farhoosh H and Zimmerman L 1980
{\it  Phys. Rev. D} {\bf 21} 317 

\bibitem{ES}
Stephani H Kramer D  MacCallum M A H Hoenselaers C and Herlt E  2003
{\it Exact Solutions of Einstein's Theory}, 
Cambridge Univ.\ Press, Cambridge, second edition.

\bibitem{bcmprd}
Bini D Cherubini C Mashhoon B 2004
{\it Phys. Rev. D} {\bf 70}  044020

\bibitem{math37} 
Mathisson M 1937
{\it Acta Phys. Polonica} {\bf 6} 167

\bibitem{papa51} 
Papapetrou A 1951
{\it Proc. Roy. Soc. London} {\bf 209} 248

\bibitem{mash74}
Mashhoon B 1974 
{\it Nature} {\bf 250} 316

\bibitem{mash88}
Mashhoon B 1988 
{\it Phys. Rev. Lett.} {\bf 61} 2639 

\bibitem{pavda}
Pravda V  and Pravdov\'a A 2000
{\it Czech. J. Phys.} {\bf 50} 333

\bibitem{podol}
Podolsk\'y J and Griffiths J B 2001
{\it Gen.\ Rel.\ Grav.\/} {\bf 33} 59

\bibitem{idcf1}
Bini D Carini P and  Jantzen R T 1997
{\it Int.\ J.\ Mod.\ Phys. D\/} {\bf 6} 1 

\bibitem{idcf2}
Bini D Carini P and  Jantzen R T 1997
{\it Int.\ J.\ Mod.\ Phys. D\/} {\bf 6} 143

\bibitem{bcgj}
Bini D Cherubini C Geralico A Jantzen R T 2004
{\it in preparation}

\bibitem{hooker}
Hooker A Greene C H Clark W 1997 
{\it Phys. Rev. A} {\bf 55} 4609 

\bibitem{cori51}
Corinaldesi E Papapetrou A 1951
{\it Proc. Roy. Soc. London} {\bf 209}

\bibitem{pir56} 
Pirani F 1956
{\it Acta Phys. Polon.} {\bf 15} 389

\bibitem{tulc59} 
Tulczyjew W 1959
{\it Acta Phys. Polon.} {\bf 18} 393

\bibitem{taub}
Taub A H 1964 
{\it J. Math. Phys.} {\bf 5} 112

\bibitem{masspin1}
Mashhoon B 1971
{\it J. Math. Phys.} {\bf 12} 1075 

\bibitem{masspin2}
Mashhoon B 1975
{\it Ann. of Phys. (NY)} {\bf 89} 254

\bibitem{dixon} 
Dixon W G 1974 
{\it Phil. Trans. Roy. Soc. London A} {\bf 277} 59

\bibitem{bdfg1}
Bini D de Felice F Geralico A 2004
{\it Class. Quantum Grav.}
{\bf 21} 5427 

\bibitem{mol}
M\o ller C 1949
{\it Commun. Dublin Inst. Adv. Studies}
Ser. A, No. 5

\end{thebibliography}
\end{document}